\documentclass[]{spie}  

\usepackage{amsmath,amsfonts,amssymb}
\usepackage{graphicx, tikz}
\usepackage[colorlinks=true, allcolors=blue]{hyperref}
\usepackage{caption}
\usepackage{mathtools}
\usepackage{subcaption}

\title{Overview of the Optical Design of the CMB-S4 Large Aperture Telescopes and Camera Optics}

\author[a]{Patricio A. Gallardo}
\author[b, c]{Kathleen Harrington}
\author[d]{Roberto Puddu}
\author[a, c, e]{Bradford Benson}
\author[a, b, c, e, f, g]{John Carlstrom}
\author[h]{Nick Emerson}
\author[a, c, e, f, g]{Jeff McMahon}
\author[a, c]{Tyler Natoli}
\author[i]{Johanna M. Nagy}
\author[j, k]{Michael D. Niemack}
\author[i]{John Ruhl}

\affil[a]{Kavli Institute for Cosmological Physics, University of Chicago, Chicago, IL, USA}
\affil[b]{High Energy Physics Division, Argonne National Laboratory, Argonne, IL, USA}
\affil[c]{Department of Astronomy and Astrophysics, University of Chicago, Chicago, IL, USA}
\affil[d]{Instituto de Astrofísica and Centro de Astro-Ingeniería, Facultad de Física, Pontificia Universidad Católica de Chile, Santiago, Chile}
\affil[e]{Fermi National Accelerator Laboratory, Batavia, IL, USA}
\affil[f]{Enrico Fermi Institute, University of Chicago, Chicago, IL, USA}
\affil[g]{Department of Physics, University of Chicago, Chicago, IL, USA}
\affil[h]{Steward Observatory, The University of Arizona, Tucson, AZ, USA}
\affil[i]{Physics Dept, Case Western Reserve University, Cleveland, OH, USA}
\affil[j]{Department of Astronomy, Cornell University, Ithaca, NY, USA}
\affil[k]{Department of Physics, Cornell University, Ithaca, NY, USA}

\author[l]{the CMB-S4 collaboration}
\authorinfo{}

\begin{document} 
\maketitle

\begin{abstract}
    CMB-S4, the next-generation CMB observatory, will deploy hundreds of thousands of detectors to enable mapping the millimeter-wavelength sky with unprecedented speed. The large aperture telescopes for CMB-S4 consist of six-meter diameter crossed Dragone designs and a five-meter diameter three-mirror anastigmat. The two-mirror crossed Dragone design requires astigmatism corrections in the refractive optics to achieve diffraction-limited performance. We present biconic lens corrections for the CMB-S4 crossed Dragone camera optics and compare these designs to the camera optics for the three mirror anastigmat, as the optical designs of the cameras for these telescopes are being prototyped.

\end{abstract}


\section{INTRODUCTION}
\label{sec:intro}  

CMB-S4, the next-generation cosmic microwave background (CMB) experiment, is designed to significantly advance the sensitivity of millimeter-wave maps of the sky, enhancing our understanding of the origin and evolution of the universe. CMB-S4 will observe the millimeter-wave sky with two surveys, using a combination of small and large aperture telescopes, which will target inflationary science, the fundamental physics of the dark universe, the spatial distribution of matter, and the time-variable millimeter-wave sky \cite{gravitational_Abazajian_2022, sciencecaseabazajian2019cmbs4}.  

The scientific targets of CMB-S4 will be met with two surveys, to be carried out using small and large aperture telescopes. One survey (called \emph{deep}), which covers $\sim 3\%$ of the sky,  targets primordial gravitational waves and inflation, while a second survey (called \emph{wide}), covers $\sim 70\%$ of the sky and targets: the dark universe, mapping matter in the cosmos and the time-variable sky \cite{sciencecaseabazajian2019cmbs4, abazajian2019cmbs4decadal,techbook_abitbol2017cmbs4}. The deep survey will be carried out with degree-resolution small aperture telescopes, in combination with a complementary 5-meter large aperture telescope with arcminute resolution, which will be used to remove contamination of the degree-scale B-modes caused by gravitational lensing of E-mode polarization, a process referred to as \emph{delensing}. The wide survey will be observed with two 6-meter telescopes with arcminute resolution. The sizes of the patches of sky for these two surveys, mapping speed considerations  and distribution of detectors at different frequencies in the focal plane motivate dedicated designs for the large aperture telescopes.

The small field to be observed by the high-resolution deep delensing survey requires a uniform focal plane sampling at the highest frequencies of the survey ($1.1\, \rm mm$), while the field to be observed by the wide-area survey allows for a more non-uniform camera distribution. The spatial uniformity needed by the delensing survey can be achieved with a five-meter three-mirror anastigmatic telescope \cite{Gallardo:24} composed of gap-free mirrors\cite{Natoli:23}, while the higher resolution of the wide survey will be carried out with a six-meter crossed Dragone telescope\cite{parshley_10.1117/12.2314073} composed of segmented panels. These two telescope designs are to be fielded with arrays of 85 optics tubes in order to populate the focal planes with $\mathcal{O}(10^5)$ detectors.

This document gives an overview and status of the optical design of the CMB-S4 telescopes and cameras building on previously presented material \cite{gallardo_22_spie_10.1117/12.2626876,Gallardo:24}, with emphasis on ongoing optimization of the crossed Dragone camera designs using biconic surfaces. Section \ref{sec:telescopes} summarizes the telescope optics and gives a brief description of their optical performance. Section \ref{sec:camoptics} discusses the optical arrangement of the cameras in the two telescope systems. Section \ref{sec:conclusion} discusses the current status and future work.

\begin{figure}
    \begin{minipage}{\textwidth}
    \centering
    $\vcenter{\hbox{\includegraphics[trim= 0.30in 0.1in 0.23in 0.5in, clip, width=0.54\textwidth]{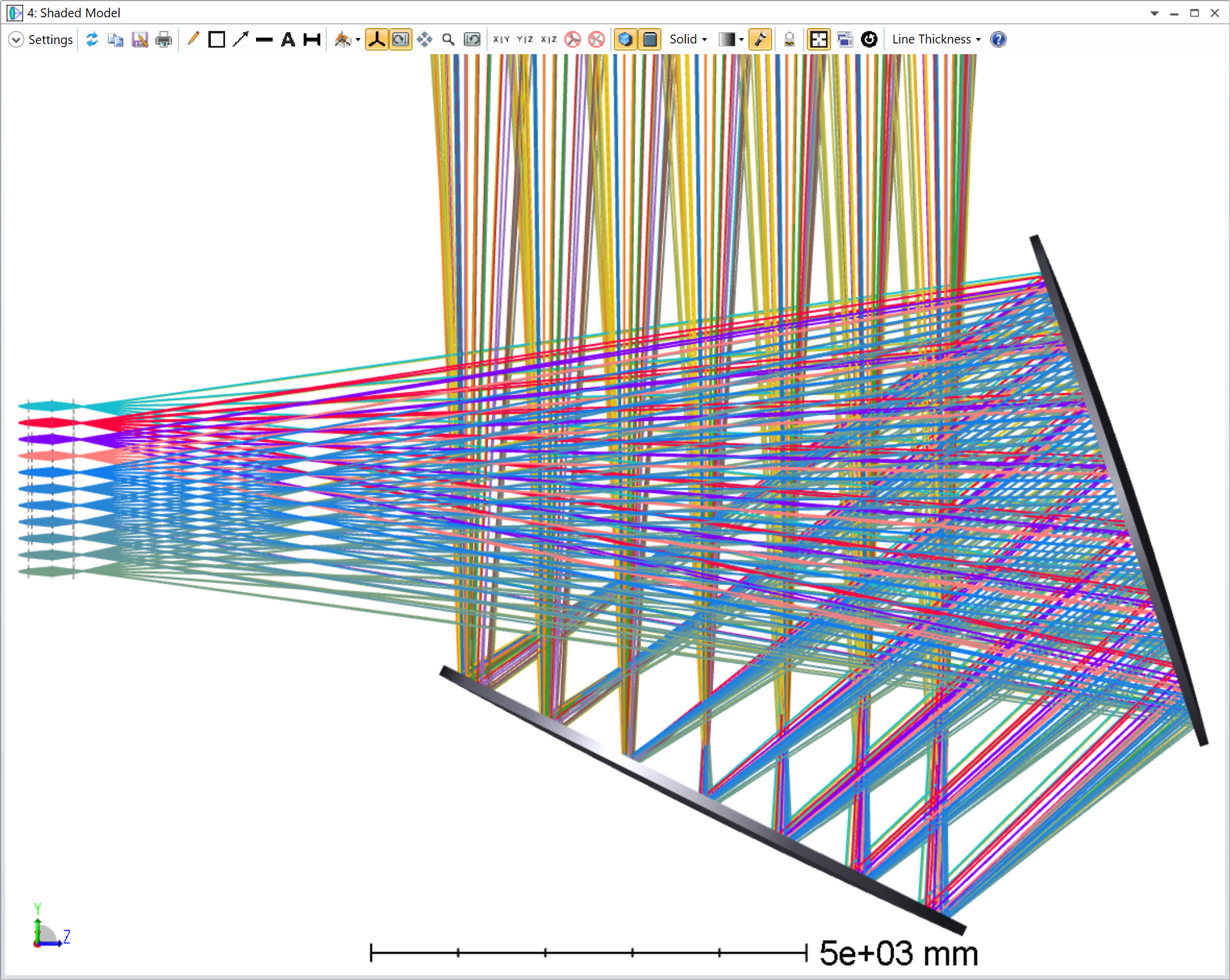}}}$   \hspace*{.0in} $\vcenter{\hbox{\includegraphics[trim= 0.20in 0.16in 0.20in 0.8in, clip, height=0.49\textheight]{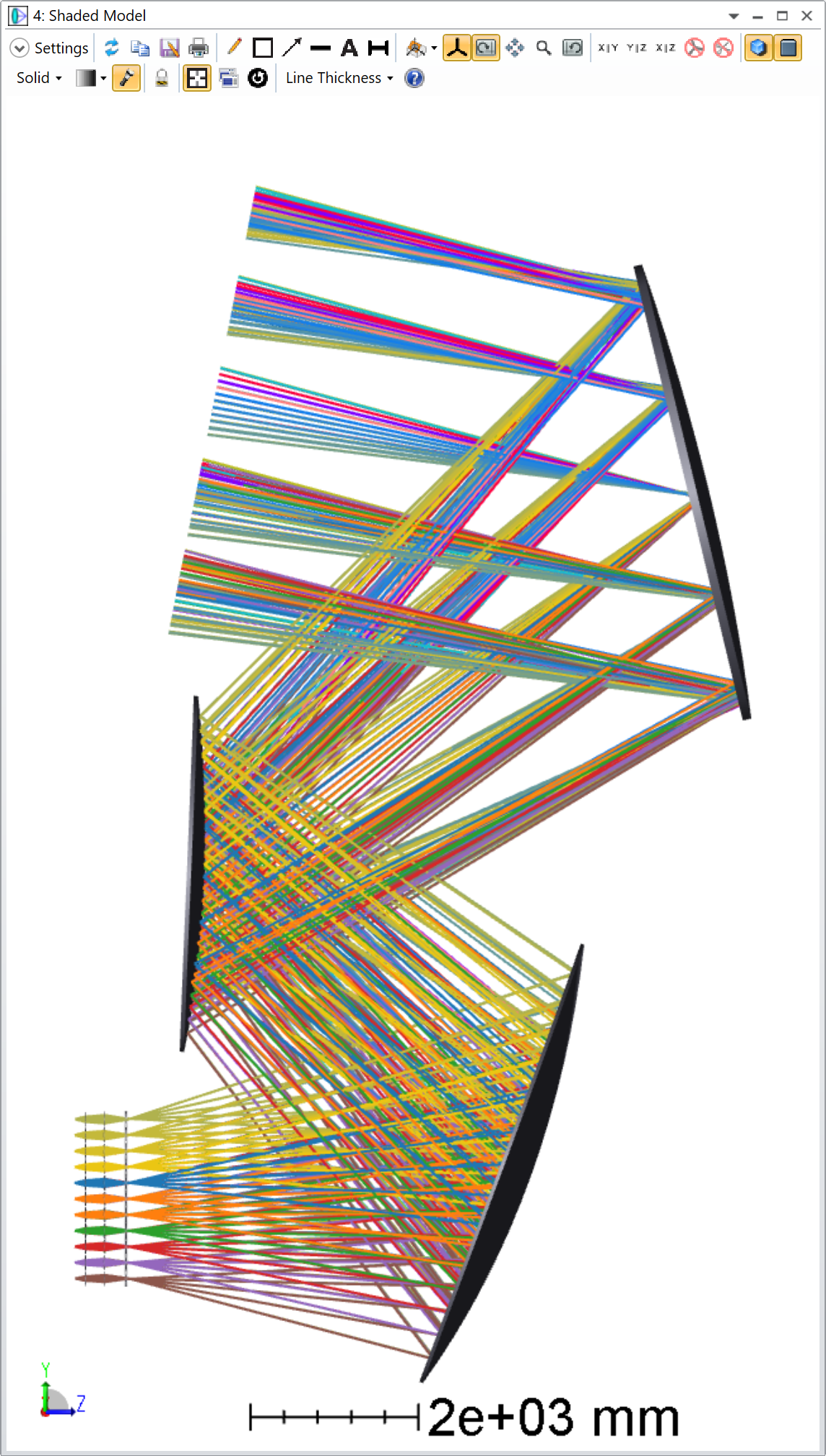}}}$
    \end{minipage}
    \centering
    \begin{tikzpicture}
        \node[inner sep=0pt] (CDcam) at (0,0) {\includegraphics*[trim= 0.3in 0.1in 0.31in 1.45in, clip, width=0.7\textwidth]{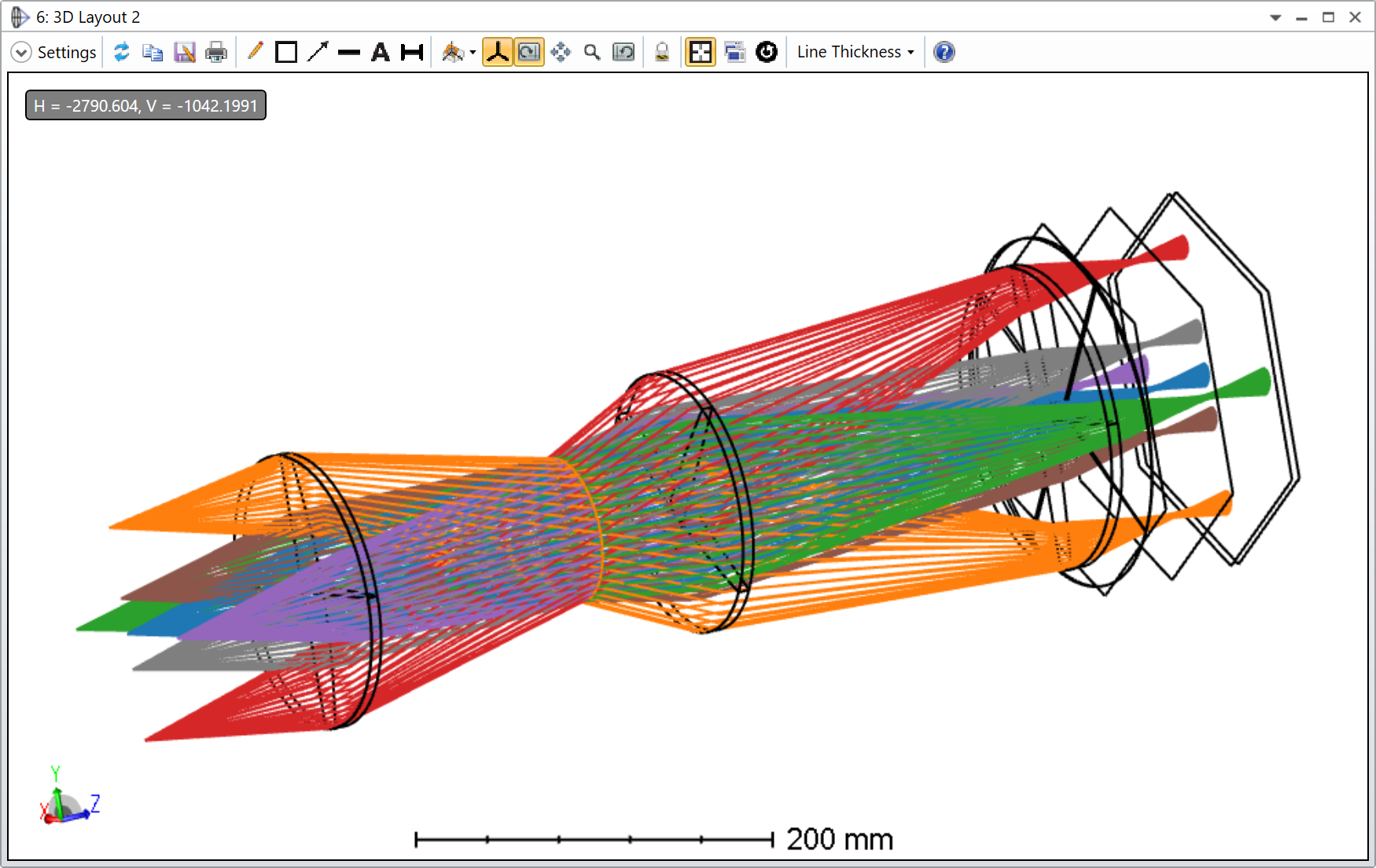}};
        \node at (-3.8,0.6) {L3};
        \node at (-0.8,-1.5) {Lyot Stop};
        \node at (-0.3,1.5) {Biconic L2};
        \node at (2.5, 2.3) {L1};
        \node at (4,-1.1) {Prism};

    \end{tikzpicture}
    \caption{\textbf{Top:} Optical layout of the two large aperture telescopes and camera concepts for CMB-S4. \textbf{Top Left:} The crossed Dragone (CD) design,  which consists of two off-axis six-meter mirrors and is able to illuminate a 7.8 degree focal plane. \textbf{Top Right:} The three-mirror anastigmatic (TMA) large aperture  telescope. The TMA consists of three off-axis free-form mirrors, with a five-meter diameter primary and is able to observe a field of view of 9 degrees in diameter. \textbf{Bottom:} The CMB-S4 crossed Dragone biconic camera concept. From left to right: focal plane, lens 3 (L3), Lyot stop, biconic lens 2 (L2), lens 1 (L1), prism, spacing for filters and window (hexagonal apertures on the right). Note the rotation of L2 which is indicated by the cross across the diameter of the lens relative to L3. The rotation angle of L2 is the same angle as the prism and is chosen such that the x-axis of the lens points towards the center of the telescope focal plane. Image shows camera 22 (see diagram in Figure \ref{fig:camgroups}). The direction of travel of light is from right to left. 
    The CMB-S4 camera concept for the three-mirror anastigmat is also a three lens system (not shown). These two camera concepts differ in the shape of lens 2 (L2) which is biconic in the crossed Dragone camera, but is radially symmetric in the TMA design.}
    \label{fig:telescopesdiagram}
\end{figure}
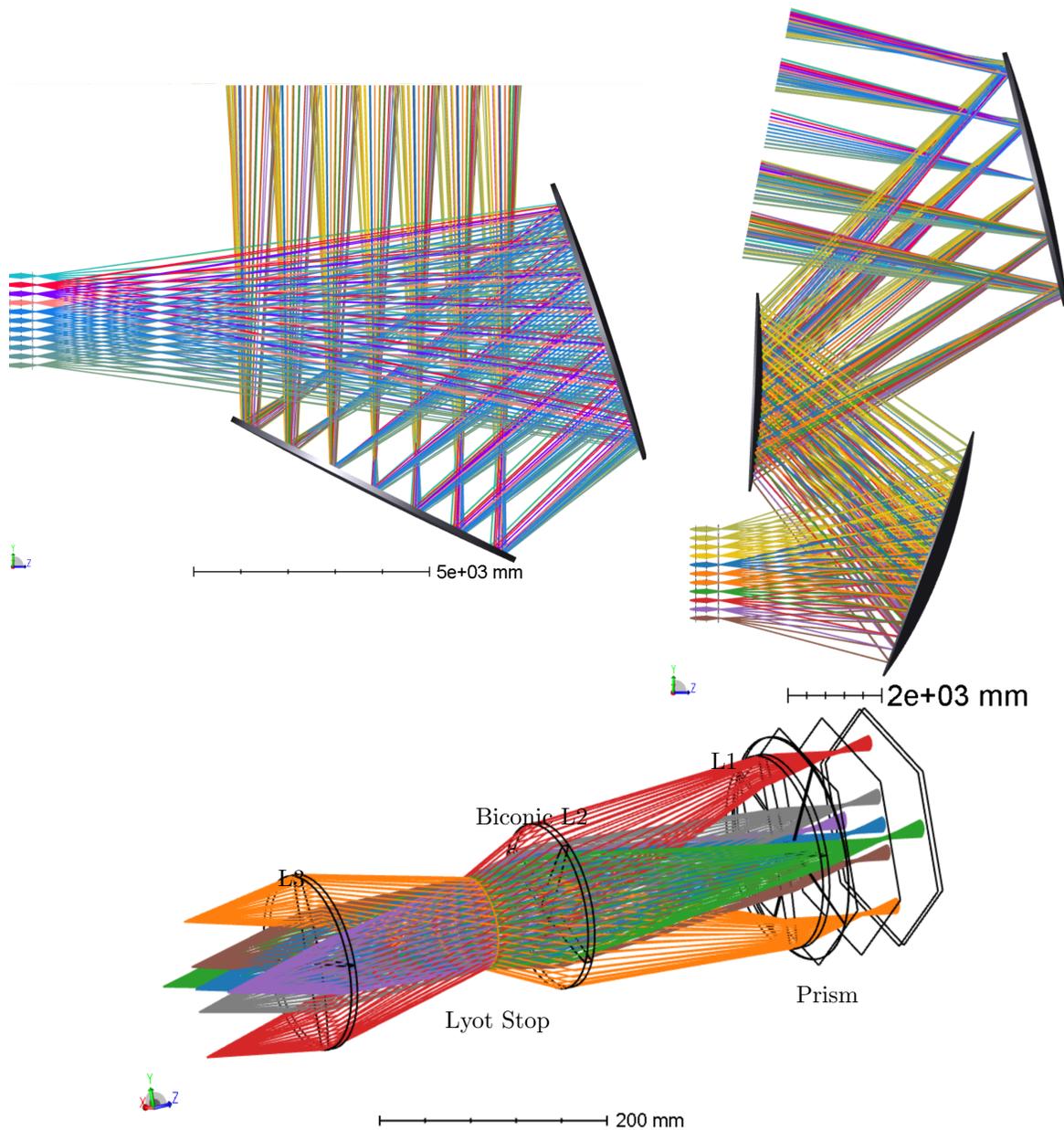

\section{Large Aperture Telescopes For CMB-S4}
\label{sec:telescopes}

Arcminute observations of the cosmic microwave background  benefit of the clean beams provided by the  unobstructed, off-axis, reflective telescope configuarion. Large aperture CMB telescopes, with a primary mirror larger than 5\,meters, like ACT and SPT, have made use of a two-mirror system in a Gregorian configuration, where the focal point of the primary is located between the two mirrors. \cite{Fowler:07,Padin:08} This configuration allows the use of multiple cameras with several thousands of detectors. However, current scientific targets demand wide, high fidelity maps of the CMB, which require larger focal planes in order to accommodate up to hundreds of thousands of detectors. The crossed Dragone configuration, in use in CCAT and the Simons Observatory\cite{parshley_10.1117/12.2314073} allows for observations with multiple cameras  over a field of view of 7.8 degrees with an aperture of 6 meters in diameter made of segmented panels. The three-mirror anastigmatic telescope enables observations of a  9-degree field with a gap-less five-meter diameter aperture with increased optical quality in the millimeter band.
Currently under-construction crossed Dragone telescopes populate the focal plane with optics tubes including large-format 40\,cm lenses \cite{2018SPIE10700E..3ED}, which are able to illuminate 39 detector wafers with $\sim 60\rm k$ detectors\cite{Zhu_2021}, while the proposed design of CMB-S4 makes use of smaller 20\,cm lenses, illuminating 85 similar detector wafers with $\sim 130\rm k$ detectors per telescope. 

The two surveys of CMB-S4 (\emph{delensing} and \emph{wide}) are planned with unique angular sizes and depths, matched to their scientific targets. The delensing survey will observe a small region constituting only of 3\% of the sky. Simultaneously, this region will be observed with several small aperture telescopes targeting inflationary science. The wide survey will observe 70\% of the sky. The differences in survey size, possible focal plane band layouts and mapping speed considerations mean that these two surveys are better performed with dedicated large aperture telescope designs, a crossed Dragone for the wide survey and a three mirror anastigmat for the deep survey. This section describes the two optical configurations of the two large aperture telescope configurations.

\subsection{Crossed Dragone}
The crossed Dragone telescope has a six-meter aperture, is composed of paneled reflectors, and the secondary mirror is of similar size to the primary, which allows to achieve large fields of view. This telescope design consists of two off-axis mirrors, which have a maximum  field of view of 7.8 degrees. The diffraction-limited field of view for this optical configuration is a function of frequency at the bands of interest \cite{parshley_10.1117/12.2314073,gallardo_22_spie_10.1117/12.2626876}. The highest image quality is achieved at the center of the field, where the highest frequency cameras, operating at $1.1\, \rm mm$ can be located. The dominant optical aberration for the off-axis fields in the  crossed Dragone design is astigmatism, which can be corrected in the camera system. A diagram of the crossed Dragone telescope with cameras is shown in Figure \ref{fig:telescopesdiagram} (top left). It has been shown that for this telescope, a three camera system where the second lens is a biconic lens yields diffraction-limited performance at a wavelength of 2 millimeters over nearly the full field of view using individually optimized cameras \cite{gallardo_22_spie_10.1117/12.2626876}. A system like this, where cameras can belong to two groups that define the lens parameters to simplify the design, is discussed in Section \ref{sec:camoptics}.

\subsection{Three Mirror Anastigmat}
The delensing survey benefits of a spatially uniform distribution of high frequency cameras over the vignetting-limited, field of view of 9 degrees. The ability to maintain high Strehl ratios over a wide field is a characteristic of the three-mirror anastigmatic design. This design is able  to cancel the dominant astigmatism of a two-mirror system by introducing a third active optical element. A system like this has been demonstrated previously\cite{Padin:18}, and optimized to match the f-number of the crossed Dragone\cite{Gallardo:24}. The three-mirror anastigmatic telescope consists of a gap-free 5-meter off-axis primary mirror, which illuminates a smaller secondary, which in turn illuminates a tertiary of roughly the same size as the primary mirror. A diagram of the three mirror anastigmat with cameras is shown in Figure \ref{fig:telescopesdiagram} (top right). This telescope achieves a diffraction-limited field of view of 9 degrees at the shortest wavelength ($\lambda = 1.1\,\rm mm$) of the survey.

\section{Camera Optics}
\label{sec:camoptics}

The camera concept for the large aperture telescopes of CMB-S4 consists of an array of 85 cameras arranged in a hexagonal tiling. This arrangement efficiently populates the focal plane, which allows us to maximize the number of on-sky detectors. The two large aperture telescope concepts have been designed with an identical target f-number, which enables the use of a shared camera concept.

All 85 cameras consist of three lenses, a Lyot stop and an alumina prism. Radially symmetric lenses (used in the three mirror anastigmat's camera and in lens 1 and 3 in the crossed Dragone camera) have a curved shape given by \begin{equation}
    \label{eq:radiallysymmetric}
    z(r) = \frac{r^2/R}{1+\sqrt{1-(1+k)(r/R)^2}},
\end{equation}
where the free parameters $R$ and $k$ are the radius of curvature and the conic constant of the surface. The asymmetric lens shape used in the second lens in the crossed Dragone camera is presented in Section \ref{subsec:cdcamoptics}.
Cryo-engineering considerations dictate the maximum allowable lens diameters for a given camera pitch (distance from one camera to the next), which are kept identical among the two  telescope cameras. Also, because the two telescope concepts for CMB-S4 have different optical aberrations, the optical prescription and positions of the lenses are different between the crossed Dragone telescope and the three mirror anastigmat, in addition, because the optical aberrations change across the focal plane of each large aperture telescope, the optimal optical  prescription varies across the telescope focal plane. The result, from a fabrication point of view, is that it is desirable to break down the design of the 85 cameras into smaller groups with identical optical prescriptions. This choice simplifies fabrication and allows a small degree of asymmetry in the crossed Dragone camera concept in order to target astigmatism corrections. A more detailed description of these two camera concepts is given below.

\begin{figure}
    \centering
    \begin{subfigure}[t]{0.48\textwidth}
    \centering
    \includegraphics[trim= 0.1in 0.15in 0.10in 0in, clip, width=\textwidth]{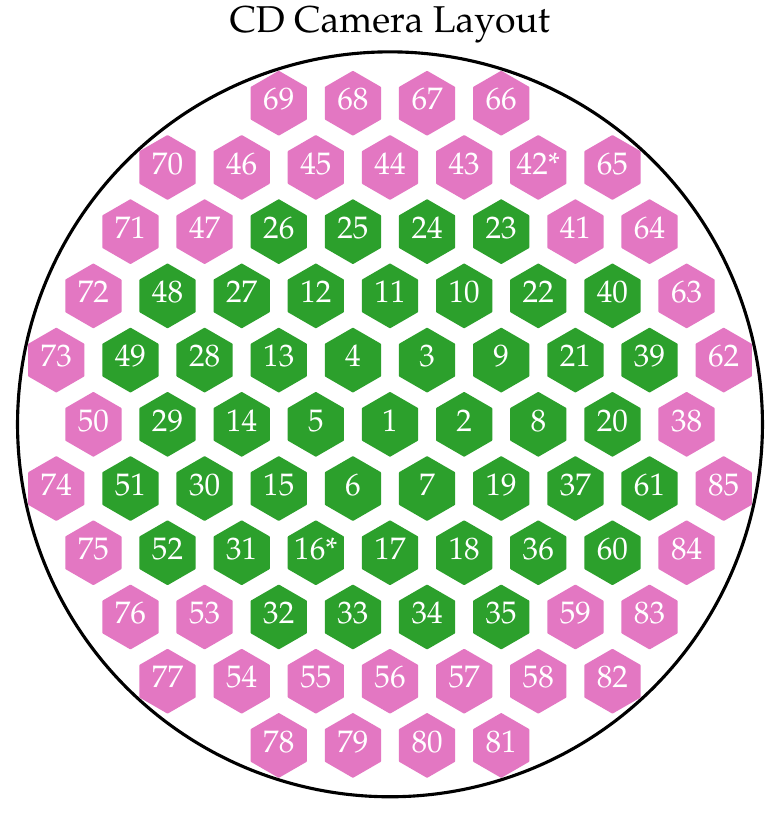}
    \end{subfigure}
    \begin{subfigure}[t]{0.48\textwidth}
    \includegraphics[trim= 0.1in 0.15in 0.10in 0in, clip, width=\textwidth]{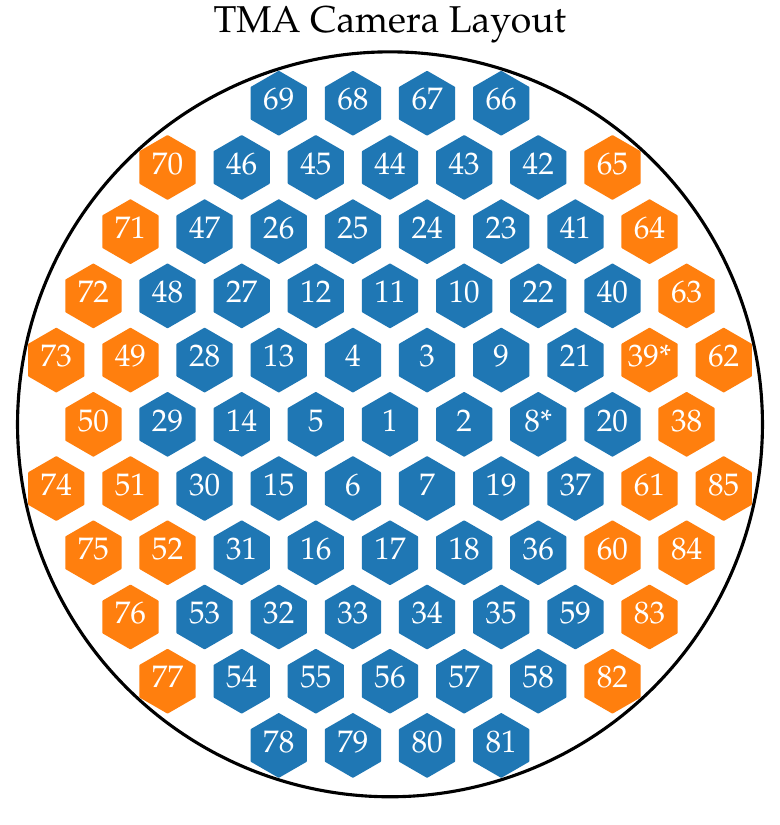}    

    \end{subfigure}
        \centering
        \begin{subfigure}[t]{0.48\textwidth}
        \centering
        \large{CD camera band distribution}
        \end{subfigure}
        \begin{subfigure}[t]{0.48\textwidth}
        \centering
        \large{TMA camera band distribution}
        \end{subfigure}

        \centering
        \begin{subfigure}[t]{0.48\textwidth}
            \centering
            \includegraphics[trim= 0.08in 0.0in 0.05in 0.2in, clip, width=\textwidth]{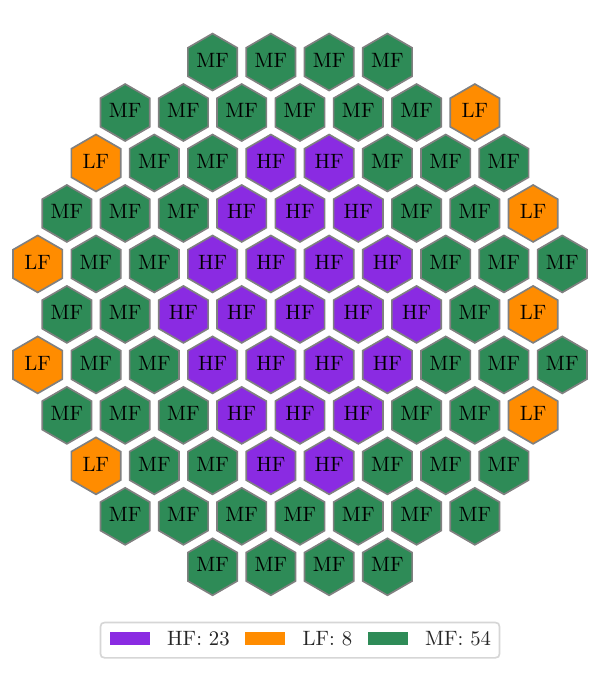}
            \end{subfigure}
            \begin{subfigure}[t]{0.48\textwidth}
            \centering
            \includegraphics[trim= 0.08in 0.0in 0.1in 0.2in, clip, width=\textwidth]{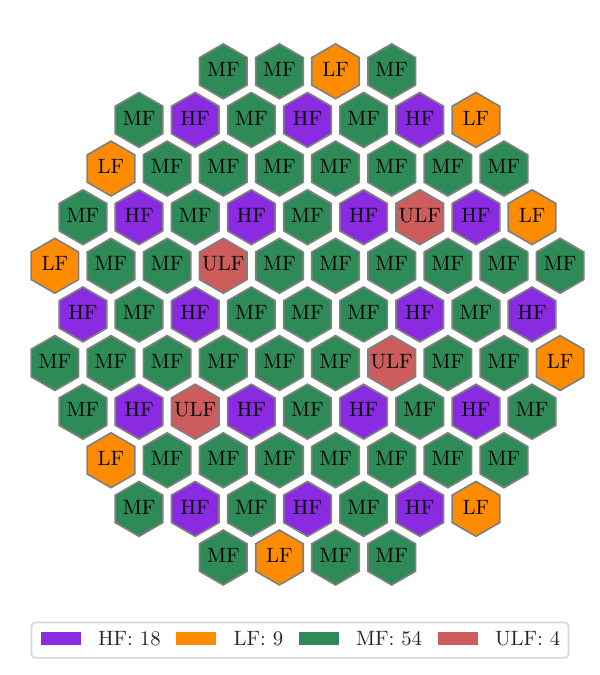}
            \end{subfigure}

        \label{fig:enter-label}

    \caption{Camera layouts for the two CMB-S4 telescope camera concepts. \textbf{Top:} Colors indicate groups of cameras with identical lens prescriptions. Top left shows the crossed Dragone cameras showing the same optical design prescription. Top right shows the three-mirror anastigmatic telescope cameras. \textbf{Bottom:} Frequency band distribution across the focal plane for the crossed Dragone (left) and three-mirror anastigmatic (right) telescope camera arrays. Bands are labeled ULF (20, 15\,GHz), LF (26, 39\,GHz), MF (92, 149\,GHz) and HF (227, 286\,GHz). The crossed Dragone telescope design yields high Strehl ratios at the center of the field; thus the central portion of the focal plane can be used at the highest frequencies ($1.1\,\rm mm$). The TMA yields consistently high Strehl ratios across the focal plane, allowing the highest frequency cameras to be spaced out across the telescope focal plane, this is beneficial for observing large angular scales on the sky.}
    \label{fig:camgroups}
\end{figure}

\subsection{Crossed Dragone Camera Optics}
\label{subsec:cdcamoptics}

As discussed in Section \ref{sec:camoptics}, the crossed Dragone camera consists of three lenses; lens 1 and lens 3 are radially symmetric and follow Equation \ref{eq:radiallysymmetric}; while the second lens has an asymmetric surface.

Astigmatism in the crossed Dragone telescope can be  corrected in the camera optics system using an asymmetric lens with two orthogonal radii of curvature. This kind of lens, often called a \emph{biconic}, follows a surface shape given by \begin{equation}z(x,y) = \frac{x^2/R_x + y^2/R_y}{1+\sqrt{1-(1+k_x)(x/R_x)^2 - (1+k_y)(y/R_y)^2}} ,
\end{equation} which contains four free parameters: the two radii of curvature ($R_x$ and $R_y$) and the two conic constants ($k_x$ and $k_y$). An additional degree of freedom can be introduced  by rotating this surface around the optical axis at a given angle. The optimal position to correct for astigmatism in a cryogenic imaging camera system is the Lyot stop. However, correcting at the Lyot stop surface would introduce  a fourth lens, which can introduce systematic effects due to potential manufacturing imperfections  and increase manufacturing costs. In order to  minimize the complexity of this system, it has been proposed and demonstrated  that the use of the second lens with a biconic surface can correct for astigmatism down to two millimeters in wavelength across the entire field if every camera is optimized individually \cite{gallardo_22_spie_10.1117/12.2626876}.

\begin{figure*}
    \centering
    \begin{subfigure}[t]{0.49\textwidth}
    \centering
    \LARGE{Crossed Dragone}
    \end{subfigure}
    \begin{subfigure}[t]{0.49\textwidth}
    \centering
    \LARGE{Three Mirror Anastigmat}
    \end{subfigure}
    \begin{subfigure}[t]{0.49\textwidth}
    \centering
    \includegraphics[trim= 0.3in 0.25in 0.31in 0.25in, clip, width=\textwidth]{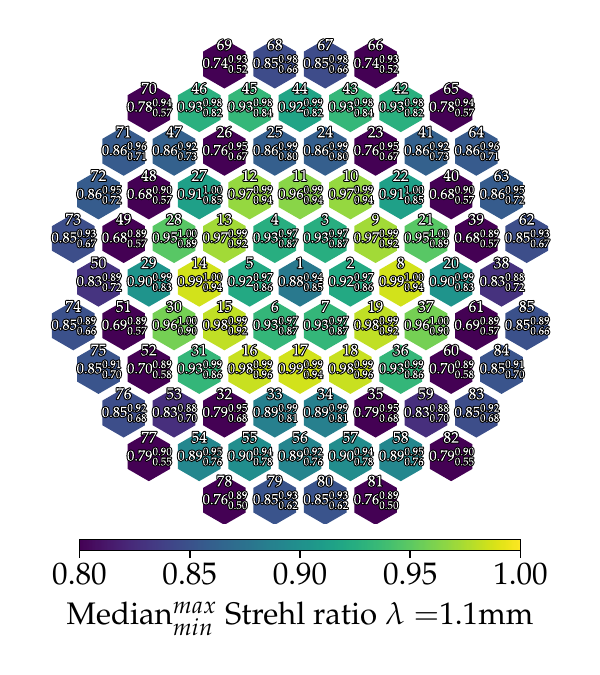}
    \end{subfigure}
    \begin{subfigure}[t]{0.49\textwidth}
    \centering
    \includegraphics[trim= 0.3in 0.25in 0.31in 0.25in, clip, width=\textwidth]{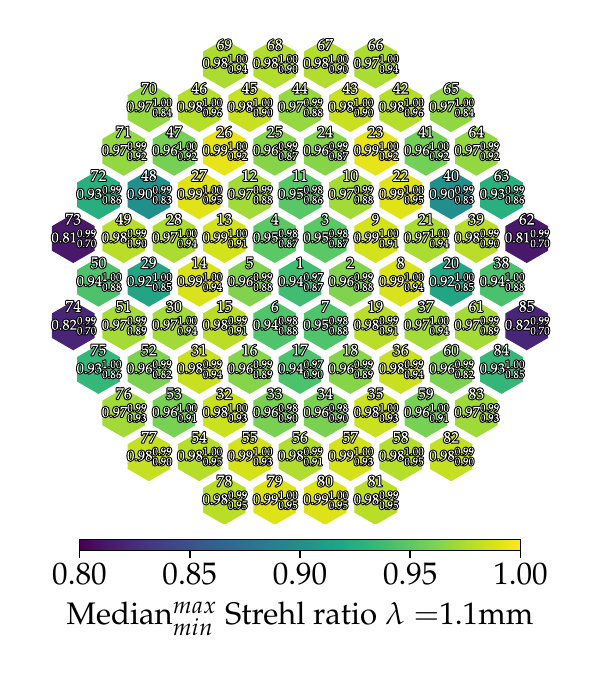}
    \end{subfigure}

    \begin{subfigure}[t]{0.49\textwidth}
    \centering
    \includegraphics[trim= 0.3in 0.25in 0.31in 0.25in, clip, width=\textwidth]{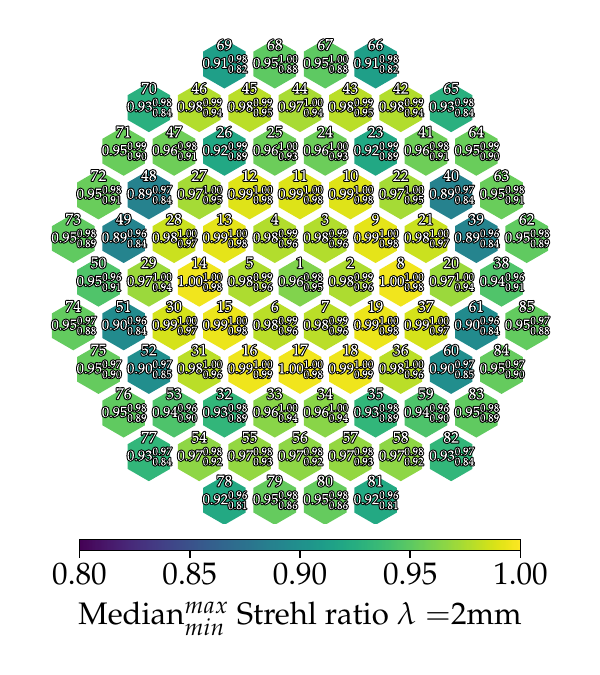}
    \end{subfigure}
    \begin{subfigure}[t]{0.49\textwidth}
    \centering
    \includegraphics[trim= 0.3in 0.25in 0.31in 0.25in, clip, width=\textwidth]{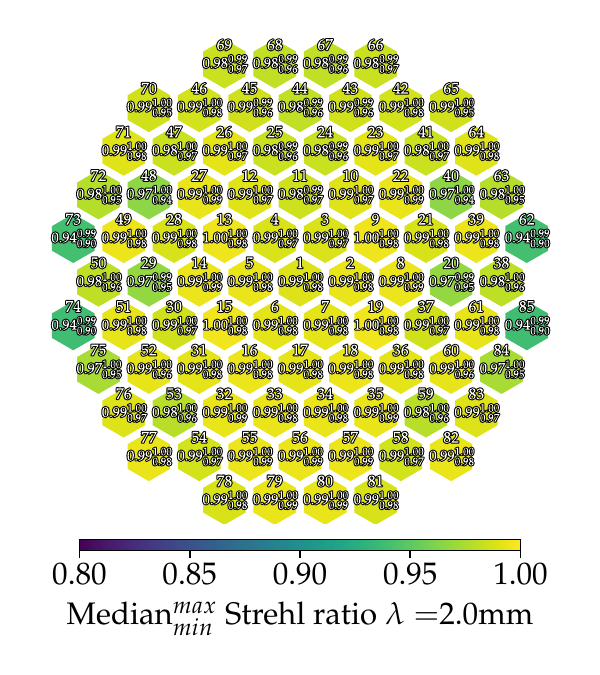}
    \end{subfigure}
    \caption{Median, maximum and minimum Strehl ratios for the two camera concepts of CMB-S4 using the tiled camera optical prescriptions shown here. Left: Crossed Dragone telescope camera concept. Right: Three Mirror Anastigmat (TMA), taken from Gallardo et al. 2024\cite{Gallardo:24}. Strehl ratios are quantified at two wavelengths: $1.1 \, \rm mm$ (top) and $2\,\rm mm$ (bottom).}
    \label{fig:imgqual}
\end{figure*}

To simplify the complexity of the optical design of the 85 cameras we explored grouping cameras using identical lens designs. Initially, we performed an extensive optimization of all cameras individually; this initial optimization searches for the best lens shapes that minimize the wavefront error at the detector focal plane. This optimization was followed by an evaluation of the entire array of cameras using an identical optical design, in order to identify which cameras can provide a diffraction-limited performance at the bands of interest. The result of this optimization procedure is captured in the two groups shown in Figure \ref{fig:camgroups} (top), which shows two groups of camera lenses that can yield the optical quality map presented in Figure \ref{fig:imgqual}. Figure \ref{fig:camgroups} (bottom) shows the camera band assignments for these telescopes. The image quality shown in Figure \ref{fig:imgqual} shows diffraction limited performance for the 85 cameras at a wavelength of $2\, \rm mm$ and the 23 high frequency $1.1 \, \rm mm $ cameras at the center of the telescope focal plane.

\subsection{Three Mirror Anastigmat Camera Optics}

The camera system for the three-mirror anastigmat is composed of three radially symmetric lenses, following a surface shape given by Equation \ref{eq:radiallysymmetric}. These lenses form an image of the primary mirror at the Lyot stop. The beam is redirected at the appropriate angle by the use of an alumina prism with a unique rotation and tilt due to the symmetry of the system. The three-mirror anastigmatic telescope focal plane  requires a unique tilt and clocking for this optical component. The optimal parameters that define the shape of the prism (rotation and tilt angle) are found numerically during the optimization in Zemax following methods presented previously \cite{gallardo_22_spie_10.1117/12.2626876,Gallardo:24}. 

The three-mirror anastigmat camera optics has been optimized following the procedure outlined in Gallardo et al. 2024 \cite{Gallardo:24}, which yields the two groups of cameras shown in Figure \ref{fig:camgroups} (top right). One group shown in blue covers the centermost cameras, while a second group corrects the cameras at the left and right extremes of the focal plane. This kind of grouping is expected as the three-mirror anastigmat has a slightly curved focal surface to the left and right of the central region of the focal surface. This focal curvature would defocus the co-planar side cameras if left uncorrected. The camera band distribution for the three mirror anastigmatic telescope cameras is shown in Figure \ref{fig:camgroups} (bottom right).

\section{Conclusion, Status and Future Work}
\label{sec:conclusion}

We have presented an overview and the current status  of the optical design of the large aperture telescopes and cameras for CMB-S4. This discussion is based on past and ongoing work towards implementing the large aperture telescopes and camera concepts. We have shown that the CMB-S4 telescope concept consists of two designs, a six-meter crossed Dragone and a five-meter three-mirror anastigmatic telescope, suited for specific target surveys and science cases. These two telescopes share a common camera concept consisting of an array of 85 three-lens cameras. We have shown that the three-mirror anastigmatic telescope can achieve diffraction-limited performance at the bands of interest using radially symmetric lenses, and the crossed Dragone needs  biconic astigmatism corrections in order to fully populate the telescope field of view with diffraction-limited cameras. A prototype of the primary mirror of the three-mirror anastimatic telescope has been built\cite{Natoli:23}, while the CMB-S4 camera optics are currently under further optimization to accommodate cryo-mechanical considerations needed for prototyping.

\acknowledgments

This material is based upon work supported by the National Science Foundation Award No. 2034402, 1935892, and 2240374. PAG acknowledges support from KICP at U. Chicago.

\bibliography{report}

\begin{thebibliography}{10}

\bibitem{gravitational_Abazajian_2022}
Abazajian, K., Addison, G.~E., Adshead, P., Ahmed, Z., Akerib, D., Ali, A.,
  Allen, S.~W., Alonso, D., Alvarez, M., Amin, M.~A., Anderson, A., Arnold,
  K.~S., Ashton, P., Baccigalupi, C., Bard, D., Barkats, D., Barron, D., Barry,
  P.~S., Bartlett, J.~G., Thakur, R.~B., Battaglia, N., Bean, R., Bebek, C.,
  Bender, A.~N., Benson, B.~A., Bianchini, F., Bischoff, C.~A., Bleem, L.,
  Bock, J.~J., Bocquet, S., Boddy, K.~K., Bond, J.~R., Borrill, J., Bouchet,
  F.~R., Brinckmann, T., Brown, M.~L., Bryan, S., Buza, V., Byrum, K., Caimapo,
  C.~H., Calabrese, E., Calafut, V., Caldwell, R., Carlstrom, J.~E., Carron,
  J., Cecil, T., Challinor, A., Chang, C.~L., Chinone, Y., Cho, H.-M.~S.,
  Cooray, A., Coulton, W., Crawford, T.~M., Crites, A., Cukierman, A.,
  Cyr-Racine, F.-Y., de~Haan, T., Delabrouille, J., Devlin, M., Valentino,
  E.~D., Dierickx, M., Dobbs, M., Duff, S., Dvorkin, C., Eimer, J., Elleflot,
  T., Errard, J., Essinger-Hileman, T., Fabbian, G., Feng, C., Ferraro, S.,
  Filippini, J.~P., Flauger, R., Flaugher, B., Fraisse, A.~A., Frolov, A.,
  Galitzki, N., Gallardo, P.~A., Galli, S., Ganga, K., Gerbino, M., Gluscevic,
  V., Goeckner-Wald, N., Green, D., Grin, D., Grohs, E., Gualtieri, R.,
  Gudmundsson, J.~E., Gullett, I., Gupta, N., Habib, S., Halpern, M.,
  Halverson, N.~W., Hanany, S., Harrington, K., Hasegawa, M., Hasselfield, M.,
  Hazumi, M., Heitmann, K., Henderson, S., Hensley, B., Hill, C., Hill, J.~C.,
  Hložek, R., Ho, S.-P.~P., Hoang, T., Holder, G., Holzapfel, W., Hood, J.,
  Hubmayr, J., Huffenberger, K.~M., Hui, H., Irwin, K., Jeong, O., Johnson,
  B.~R., Jones, W.~C., Kang, J.~H., Karkare, K.~S., Katayama, N., Keskitalo,
  R., Kisner, T., Knox, L., Koopman, B.~J., Kosowsky, A., Kovac, J., Kovetz,
  E.~D., Kuhlmann, S., lin Kuo, C., Kusaka, A., Lähteenmäki, A., Lawrence,
  C.~R., Lee, A.~T., Lewis, A., Li, D., Linder, E., Loverde, M., Lowitz, A.,
  Lubin, P., Madhavacheril, M.~S., Mantz, A., Marques, G., Matsuda, F.,
  Mauskopf, P., McCarrick, H., McMahon, J., Meerburg, P.~D., Melin, J.-B.,
  Menanteau, F., Meyers, J., Millea, M., Mohr, J., Moncelsi, L., Monzani, M.,
  Mroczkowski, T., Mukherjee, S., Nagy, J., Namikawa, T., Nati, F., Natoli, T.,
  Newburgh, L., Niemack, M.~D., Nishino, H., Nord, B., Novosad, V., O’Brient,
  R., Padin, S., Palladino, S., Partridge, B., Petravick, D., Pierpaoli, E.,
  Pogosian, L., Prabhu, K., Pryke, C., Puglisi, G., Racine, B., Rahlin, A.,
  Rao, M.~S., Raveri, M., Reichardt, C.~L., Remazeilles, M., Rocha, G., Roe,
  N.~A., Roy, A., Ruhl, J.~E., Salatino, M., Saliwanchik, B., Schaan, E.,
  Schillaci, A., Schmitt, B., Schmittfull, M.~M., Scott, D., Sehgal, N.,
  Shandera, S., Sherwin, B.~D., Shirokoff, E., Simon, S.~M., Slosar, A.,
  Spergel, D., Germaine, T.~S., Staggs, S.~T., Stark, A., Starkman, G.~D.,
  Stompor, R., Stoughton, C., Suzuki, A., Tajima, O., Teply, G.~P., Thompson,
  K., Thorne, B., Timbie, P., Tomasi, M., Tristram, M., Tucker, G., Umiltà,
  C., van Engelen, A., Vavagiakis, E.~M., Vieira, J.~D., Vieregg, A.~G.,
  Wagoner, K., Wallisch, B., Wang, G., Watson, S., Westbrook, B., Whitehorn,
  N., Wollack, E.~J., Wu, W. L.~K., Xu, Z., Yang, H. Y.~E., Yasini, S.,
  Yefremenko, V.~G., Yoon, K.~W., Young, E., Yu, C., Zonca, A., and
  Collaboration, T. C.-S., ``Cmb-s4: Forecasting constraints on primordial
  gravitational waves,'' {\em The Astrophysical Journal}~{\bf 926},  54 (feb
  2022).

\bibitem{sciencecaseabazajian2019cmbs4}
Abazajian, K., Addison, G., Adshead, P., Ahmed, Z., Allen, S.~W., Alonso, D.,
  Alvarez, M., Anderson, A., Arnold, K.~S., Baccigalupi, C., Bailey, K.,
  Barkats, D., Barron, D., Barry, P.~S., Bartlett, J.~G., Thakur, R.~B.,
  Battaglia, N., Baxter, E., Bean, R., Bebek, C., Bender, A.~N., Benson, B.~A.,
  Berger, E., Bhimani, S., Bischoff, C.~A., Bleem, L., Bocquet, S., Boddy, K.,
  Bonato, M., Bond, J.~R., Borrill, J., Bouchet, F.~R., Brown, M.~L., Bryan,
  S., Burkhart, B., Buza, V., Byrum, K., Calabrese, E., Calafut, V., Caldwell,
  R., Carlstrom, J.~E., Carron, J., Cecil, T., Challinor, A., Chang, C.~L.,
  Chinone, Y., Cho, H.-M.~S., Cooray, A., Crawford, T.~M., Crites, A.,
  Cukierman, A., Cyr-Racine, F.-Y., de~Haan, T., de~Zotti, G., Delabrouille,
  J., Demarteau, M., Devlin, M., Valentino, E.~D., Dobbs, M., Duff, S.,
  Duivenvoorden, A., Dvorkin, C., Edwards, W., Eimer, J., Errard, J.,
  Essinger-Hileman, T., Fabbian, G., Feng, C., Ferraro, S., Filippini, J.~P.,
  Flauger, R., Flaugher, B., Fraisse, A.~A., Frolov, A., Galitzki, N., Galli,
  S., Ganga, K., Gerbino, M., Gilchriese, M., Gluscevic, V., Green, D., Grin,
  D., Grohs, E., Gualtieri, R., Guarino, V., Gudmundsson, J.~E., Habib, S.,
  Haller, G., Halpern, M., Halverson, N.~W., Hanany, S., Harrington, K.,
  Hasegawa, M., Hasselfield, M., Hazumi, M., Heitmann, K., Henderson, S.,
  Henning, J.~W., Hill, J.~C., Hlozek, R., Holder, G., Holzapfel, W., Hubmayr,
  J., Huffenberger, K.~M., Huffer, M., Hui, H., Irwin, K., Johnson, B.~R.,
  Johnstone, D., Jones, W.~C., Karkare, K., Katayama, N., Kerby, J., Kernovsky,
  S., Keskitalo, R., Kisner, T., Knox, L., Kosowsky, A., Kovac, J., Kovetz,
  E.~D., Kuhlmann, S., lin Kuo, C., Kurita, N., Kusaka, A., Lahteenmaki, A.,
  Lawrence, C.~R., Lee, A.~T., Lewis, A., Li, D., Linder, E., Loverde, M.,
  Lowitz, A., Madhavacheril, M.~S., Mantz, A., Matsuda, F., Mauskopf, P.,
  McMahon, J., McQuinn, M., Meerburg, P.~D., Melin, J.-B., Meyers, J., Millea,
  M., Mohr, J., Moncelsi, L., Mroczkowski, T., Mukherjee, S., Münchmeyer, M.,
  Nagai, D., Nagy, J., Namikawa, T., Nati, F., Natoli, T., Negrello, M.,
  Newburgh, L., Niemack, M.~D., Nishino, H., Nordby, M., Novosad, V., O'Connor,
  P., Obied, G., Padin, S., Pandey, S., Partridge, B., Pierpaoli, E., Pogosian,
  L., Pryke, C., Puglisi, G., Racine, B., Raghunathan, S., Rahlin, A.,
  Rajagopalan, S., Raveri, M., Reichanadter, M., Reichardt, C.~L., Remazeilles,
  M., Rocha, G., Roe, N.~A., Roy, A., Ruhl, J., Salatino, M., Saliwanchik, B.,
  Schaan, E., Schillaci, A., Schmittfull, M.~M., Scott, D., Sehgal, N.,
  Shandera, S., Sheehy, C., Sherwin, B.~D., Shirokoff, E., Simon, S.~M.,
  Slosar, A., Somerville, R., Spergel, D., Staggs, S.~T., Stark, A., Stompor,
  R., Story, K.~T., Stoughton, C., Suzuki, A., Tajima, O., Teply, G.~P.,
  Thompson, K., Timbie, P., Tomasi, M., Treu, J.~I., Tristram, M., Tucker, G.,
  Umiltà, C., van Engelen, A., Vieira, J.~D., Vieregg, A.~G., Vogelsberger,
  M., Wang, G., Watson, S., White, M., Whitehorn, N., Wollack, E.~J., Wu, W.
  L.~K., Xu, Z., Yasini, S., Yeck, J., Yoon, K.~W., Young, E., and Zonca, A.,
  ``Cmb-s4 science case, reference design, and project plan,'' (2019).

\bibitem{abazajian2019cmbs4decadal}
Abazajian, K., Addison, G., Adshead, P., Ahmed, Z., Allen, S.~W., Alonso, D.,
  Alvarez, M., Amin, M.~A., Anderson, A., Arnold, K.~S., Baccigalupi, C.,
  Bailey, K., Barkats, D., Barron, D., Barry, P.~S., Bartlett, J.~G., Thakur,
  R.~B., Battaglia, N., Baxter, E., Bean, R., Bebek, C., Bender, A.~N., Benson,
  B.~A., Berger, E., Bhimani, S., Bischoff, C.~A., Bleem, L., Bock, J.~J.,
  Bocquet, S., Boddy, K., Bonato, M., Bond, J.~R., Borrill, J., Bouchet, F.~R.,
  Brown, M.~L., Bryan, S., Burkhart, B., Buza, V., Byrum, K., Calabrese, E.,
  Calafut, V., Caldwell, R., Carlstrom, J.~E., Carron, J., Cecil, T.,
  Challinor, A., Chang, C.~L., Chinone, Y., Cho, H.-M.~S., Cooray, A.,
  Crawford, T.~M., Crites, A., Cukierman, A., Cyr-Racine, F.-Y., de~Haan, T.,
  de~Zotti, G., Delabrouille, J., Demarteau, M., Devlin, M., Valentino, E.~D.,
  Dobbs, M., Duff, S., Duivenvoorden, A., Dvorkin, C., Edwards, W., Eimer, J.,
  Errard, J., Essinger-Hileman, T., Fabbian, G., Feng, C., Ferraro, S.,
  Filippini, J.~P., Flauger, R., Flaugher, B., Fraisse, A.~A., Frolov, A.,
  Galitzki, N., Galli, S., Ganga, K., Gerbino, M., Gilchriese, M., Gluscevic,
  V., Green, D., Grin, D., Grohs, E., Gualtieri, R., Guarino, V., Gudmundsson,
  J.~E., Habib, S., Haller, G., Halpern, M., Halverson, N.~W., Hanany, S.,
  Harrington, K., Hasegawa, M., Hasselfield, M., Hazumi, M., Heitmann, K.,
  Henderson, S., Henning, J.~W., Hill, J.~C., Hlozek, R., Holder, G.,
  Holzapfel, W., Hubmayr, J., Huffenberger, K.~M., Huffer, M., Hui, H., Irwin,
  K., Johnson, B.~R., Johnstone, D., Jones, W.~C., Karkare, K., Katayama, N.,
  Kerby, J., Kernovsky, S., Keskitalo, R., Kisner, T., Knox, L., Kosowsky, A.,
  Kovac, J., Kovetz, E.~D., Kuhlmann, S., lin Kuo, C., Kurita, N., Kusaka, A.,
  Lahteenmaki, A., Lawrence, C.~R., Lee, A.~T., Lewis, A., Li, D., Linder, E.,
  Loverde, M., Lowitz, A., Madhavacheril, M.~S., Mantz, A., Matsuda, F.,
  Mauskopf, P., McMahon, J., Meerburg, P.~D., Melin, J.-B., Meyers, J., Millea,
  M., Mohr, J., Moncelsi, L., Mroczkowski, T., Mukherjee, S., Münchmeyer, M.,
  Nagai, D., Nagy, J., Namikawa, T., Nati, F., Natoli, T., Negrello, M.,
  Newburgh, L., Niemack, M.~D., Nishino, H., Nordby, M., Novosad, V., O'Connor,
  P., Obied, G., Padin, S., Pandey, S., Partridge, B., Pierpaoli, E., Pogosian,
  L., Pryke, C., Puglisi, G., Racine, B., Raghunathan, S., Rahlin, A.,
  Rajagopalan, S., Raveri, M., Reichanadter, M., Reichardt, C.~L., Remazeilles,
  M., Rocha, G., Roe, N.~A., Roy, A., Ruhl, J., Salatino, M., Saliwanchik, B.,
  Schaan, E., Schillaci, A., Schmittfull, M.~M., Scott, D., Sehgal, N.,
  Shandera, S., Sheehy, C., Sherwin, B.~D., Shirokoff, E., Simon, S.~M.,
  Slosar, A., Somerville, R., Staggs, S.~T., Stark, A., Stompor, R., Story,
  K.~T., Stoughton, C., Suzuki, A., Tajima, O., Teply, G.~P., Thompson, K.,
  Timbie, P., Tomasi, M., Treu, J.~I., Tristram, M., Tucker, G., Umiltà, C.,
  van Engelen, A., Vieira, J.~D., Vieregg, A.~G., Vogelsberger, M., Wang, G.,
  Watson, S., White, M., Whitehorn, N., Wollack, E.~J., Wu, W. L.~K., Xu, Z.,
  Yasini, S., Yeck, J., Yoon, K.~W., Young, E., and Zonca, A., ``Cmb-s4 decadal
  survey apc white paper,'' (2019).

\bibitem{techbook_abitbol2017cmbs4}
Abitbol, M.~H., Ahmed, Z., Barron, D., Thakur, R.~B., Bender, A.~N., Benson,
  B.~A., Bischoff, C.~A., Bryan, S.~A., Carlstrom, J.~E., Chang, C.~L., Chuss,
  D.~T., Crowley, K.~T., Cukierman, A., de~Haan, T., Dobbs, M.,
  Essinger-Hileman, T., Filippini, J.~P., Ganga, K., Gudmundsson, J.~E.,
  Halverson, N.~W., Hanany, S., Henderson, S.~W., Hill, C.~A., Ho, S.-P.~P.,
  Hubmayr, J., Irwin, K., Jeong, O., Johnson, B.~R., Kernasovskiy, S.~A.,
  Kovac, J.~M., Kusaka, A., Lee, A.~T., Maria, S., Mauskopf, P., McMahon,
  J.~J., Moncelsi, L., Nadolski, A.~W., Nagy, J.~M., Niemack, M.~D., O'Brient,
  R.~C., Padin, S., Parshley, S.~C., Pryke, C., Roe, N.~A., Rostem, K., Ruhl,
  J., Simon, S.~M., Staggs, S.~T., Suzuki, A., Switzer, E.~R., Tajima, O.,
  Thompson, K.~L., Timbie, P., Tucker, G.~S., Vieira, J.~D., Vieregg, A.~G.,
  Westbrook, B., Wollack, E.~J., Yoon, K.~W., Young, K.~S., and Young, E.~Y.,
  ``Cmb-s4 technology book, first edition,'' (2017).

\bibitem{Gallardo:24}
Gallardo, P.~A., Puddu, R., Harrington, K., Benson, B., Carlstrom, J.~E.,
  Dicker, S.~R., Emerson, N., Gudmundsson, J.~E., Limon, M., McMahon, J., Nagy,
  J.~M., Natoli, T., Niemack, M.~D., Padin, S., Ruhl, J., Simon, S.~M., and
  Collaboration, T. C.-S., ``Freeform three-mirror anastigmatic large-aperture
  telescope andreceiver optics for cmb-s4,'' {\em Appl. Opt.}~{\bf 63},
  310--321 (Jan 2024).

\bibitem{Natoli:23}
Natoli, T., Benson, B., Carlstrom, J., Chauvin, E., Clavel, B., Emerson, N.,
  Gallardo, P., Niemack, M., Padin, S., Schwab, K., Stenvers, L., and Zivick,
  J., ``Fabrication of a monolithic 5m aluminum reflector for
  millimeter-wavelength observations of the cosmic microwave background,'' {\em
  Appl. Opt.}~{\bf 62},  4747--4752 (Jun 2023).

\bibitem{parshley_10.1117/12.2314073}
Parshley, S.~C., Niemack, M., Hills, R., Dicker, S.~R., D{\"u}nner, R., Erler,
  J., Gallardo, P.~A., Gudmundsson, J.~E., Herter, T., Koopman, B.~J., Limon,
  M., Matsuda, F.~T., Mauskopf, P., Riechers, D.~A., Stacey, G.~J., and
  Vavagiakis, E.~M., ``{The optical design of the six-meter CCAT-prime and
  Simons Observatory telescopes},'' in [{\em Ground-based and Airborne
  Telescopes VII}{\nolinebreak\hspace{0.1em}]},  Marshall, H.~K. and
  Spyromilio, J., eds.,  {\bf 10700},  1070041, International Society for
  Optics and Photonics, SPIE (2018).

\bibitem{gallardo_22_spie_10.1117/12.2626876}
Gallardo, P.~A., Benson, B., Carlstrom, J., Dicker, S.~R., Emerson, N.,
  Gudmundsson, J.~E., Hills, R., Limon, M., McMahon, J., Niemack, M.~D., Nagy,
  J.~M., Padin, S., Ruhl, J., and Simon, S.~M., ``{Optical design concept of
  the CMB-S4 large-aperture telescopes and cameras},'' in [{\em Millimeter,
  Submillimeter, and Far-Infrared Detectors and Instrumentation for Astronomy
  XI}{\nolinebreak\hspace{0.1em}]},  Zmuidzinas, J. and Gao, J.-R., eds.,  {\bf
  12190},  121900C, International Society for Optics and Photonics, SPIE
  (2022).

\bibitem{Fowler:07}
Fowler, J.~W., Niemack, M.~D., Dicker, S.~R., Aboobaker, A.~M., Ade, P. A.~R.,
  Battistelli, E.~S., Devlin, M.~J., Fisher, R.~P., Halpern, M., Hargrave,
  P.~C., Hincks, A.~D., Kaul, M., Klein, J., Lau, J.~M., Limon, M., Marriage,
  T.~A., Mauskopf, P.~D., Page, L., Staggs, S.~T., Swetz, D.~S., Switzer,
  E.~R., Thornton, R.~J., and Tucker, C.~E., ``Optical design of the atacama
  cosmology telescope and the millimeter bolometric array camera,'' {\em Appl.
  Opt.}~{\bf 46},  3444--3454 (Jun 2007).

\bibitem{Padin:08}
Padin, S., Staniszewski, Z., Keisler, R., Joy, M., Stark, A.~A., Ade, P. A.~R.,
  Aird, K.~A., Benson, B.~A., Bleem, L.~E., Carlstrom, J.~E., Chang, C.~L.,
  Crawford, T.~M., Crites, A.~T., Dobbs, M.~A., Halverson, N.~W., Heimsath, S.,
  Hills, R.~E., Holzapfel, W.~L., Lawrie, C., Lee, A.~T., Leitch, E.~M., Leong,
  J., Lu, W., Lueker, M., McMahon, J.~J., Meyer, S.~S., Mohr, J.~J., Montroy,
  T.~E., Plagge, T., Pryke, C., Ruhl, J.~E., Schaffer, K.~K., Shirokoff, E.,
  Spieler, H.~G., and Vieira, J.~D., ``South pole telescope optics,'' {\em
  Appl. Opt.}~{\bf 47},  4418--4428 (Aug 2008).

\bibitem{2018SPIE10700E..3ED}
{Dicker}, S.~R., {Gallardo}, P.~A., {Gudmundsson}, J.~E., {Mauskopf}, P.~D.,
  {Ali}, A., {Ashton}, P.~C., {Coppi}, G., {Devlin}, M.~J., {Galitzki}, N.,
  {Ho}, S.~P., {Hill}, C.~A., {Hubmayr}, J., {Keating}, B., {Lee}, A.~T.,
  {Limon}, M., {Matsuda}, F., {McMahon}, J., {Niemack}, M.~D.,
  {Orlowski-Scherer}, J.~L., {Piccirillo}, L., {Salatino}, M., {Simon}, S.~M.,
  {Staggs}, S.~T., {Thornton}, R., {Ullom}, J.~N., {Vavagiakis}, E.~M.,
  {Wollack}, E.~J., {Xu}, Z., and {Zhu}, N., ``{Cold optical design for the
  large aperture Simons' Observatory telescope},'' in [{\em Ground-based and
  Airborne Telescopes VII}{\nolinebreak\hspace{0.1em}]},  {Marshall}, H.~K. and
  {Spyromilio}, J., eds., {\em Society of Photo-Optical Instrumentation
  Engineers (SPIE) Conference Series} {\bf 10700},  107003E (jul 2018).

\bibitem{Zhu_2021}
Zhu, N., Bhandarkar, T., Coppi, G., Kofman, A.~M., Orlowski-Scherer, J.~L., Xu,
  Z., Adachi, S., Ade, P., Aiola, S., Austermann, J., Bazarko, A.~O., Beall,
  J.~A., Bhimani, S., Bond, J.~R., Chesmore, G.~E., Choi, S.~K., Connors, J.,
  Cothard, N.~F., Devlin, M., Dicker, S., Dober, B., Duell, C.~J., Duff, S.~M.,
  Dünner, R., Fabbian, G., Galitzki, N., Gallardo, P.~A., Golec, J.~E.,
  Haridas, S.~K., Harrington, K., Healy, E., Ho, S.-P.~P., Huber, Z.~B.,
  Hubmayr, J., Iuliano, J., Johnson, B.~R., Keating, B., Kiuchi, K., Koopman,
  B.~J., Lashner, J., Lee, A.~T., Li, Y., Limon, M., Link, M., Lucas, T.~J.,
  McCarrick, H., Moore, J., Nati, F., Newburgh, L.~B., Niemack, M.~D.,
  Pierpaoli, E., Randall, M.~J., Sarmiento, K.~P., Saunders, L.~J., Seibert,
  J., Sierra, C., Sonka, R., Spisak, J., Sutariya, S., Tajima, O., Teply,
  G.~P., Thornton, R.~J., Tsan, T., Tucker, C., Ullom, J., Vavagiakis, E.~M.,
  Vissers, M.~R., Walker, S., Westbrook, B., Wollack, E.~J., and Zannoni, M.,
  ``The simons observatory large aperture telescope receiver,'' {\em The
  Astrophysical Journal Supplement Series}~{\bf 256},  23 (sep 2021).

\bibitem{Padin:18}
Padin, S., ``Three-mirror anastigmat for cosmic microwave background
  observations,'' {\em Appl. Opt.}~{\bf 57},  2314--2326 (Mar 2018).

\end{thebibliography}
\bibliographystyle{spiebib} 

\end{document}